# Comparison of HRV Indices of ECG and BCG Signals

Kavya Remesh[1] and Job Chunkath[2]

[1] *Dept. of ECE, Dadi Institute of Engineering and Technology, Anakapalle, Andhra Pradesh*
[2] *Dept. of ECE, Government Engineering College, Thrissur, Kerala*

**Abstract-** Electrocardiography (ECG) plays a significant role in diagnosing heart-related issues, it provides, accurate, fast, and dependable insights into crucial parameters like QRS complex duration, the R-R interval, and the occurrence, amplitude, and duration of P, R, and T waves. However, utilizing ECG for prolonged monitoring poses challenges as it necessitates connecting multiple electrodes to the patient's body. This can be discomforting and disruptive, hampering the attainment of uninterrupted recordings. Ballistocardiography (BCG) emerges as a promising substitute for ECG, presenting a non-invasive technique for recording the heart's mechanical activity. BCG signals can be captured using sensors positioned beneath the bed, thereby providing enhanced comfort and convenience for long-term monitoring of the subject. In a recent study, researchers compared the heart rate variability (HRV) indices derived from simultaneously acquired ECG and BCG signals. Encouragingly, the BCG signal yielded satisfactory results similar to those obtained from ECG, implying that BCG holds potential as a viable alternative for prolonged monitoring. The findings of this study carry substantial implications for the advancement of innovative, non-invasive methods in monitoring heart health. BCG showcases the ability to offer a more comfortable and convenient alternative to ECG while retaining its capacity to deliver accurate and reliable cardiac information concerning a patient's condition.

## I. INTRODUCTION

Electrocardiography (ECG) is a well-established technique for diagnosing cardiac function, involving the transcutaneous interpretation of the heart's electrical activity through electrodes placed on the skin and recorded externally [1,2]. A standard ECG waveform consists of five waves, namely P, Q, R, S, and T [3].

Ballistocardiography (BCG) emerges as an inconspicuous substitute for ECG, capturing body movements resulting from shifts in the body's center of mass caused by arterial blood displacement and cardiac contraction [4]. In short, BCG generates a graphical representation of repetitive bodily motion resulting from the forceful ejection of blood into the major vessels with each heartbeat. This vital sign consists of eight fiducial points (G, H, I, J, K, L, M, and N), falling within the frequency range of 1-20 Hz [5,6,7].

Recent advancements in BCG instrumentation have bolstered its viability as an ECG alternative. Cost-effective and compact measurement tools, such as piezoelectric sensors (EMFi), static-charge-sensitive beds, force plates, and modified commercial weighing scales, have been developed for BCG [5, 8]. The primary advantage of BCG over ECG lies in the absence of electrodes, textiles, or similar devices that need to be affixed to the patient's body [9]. Consequently, ballistocardiography systems are well-suited for continuous monitoring of cardiopulmonary activity during sleep and over extended periods [6]. BCG offers superior patient safety and comfort while providing more comprehensive information about the heart and cardiovascular diseases due to its larger number of fiducial points [7].

Heart rate variability (HRV) represents a significant non-invasive tool for assessing cardiac autonomic activity, typically derived from beat-to-beat (RR) interval series extracted from electrocardiograms (ECGs). Accurate HRV analysis relies on high-quality ECG/BCG recordings [10].

This study primarily focuses on comparing HRV indices (time domain: SDNN, RMSSD, pNN50; frequency domain: VLF, LF, and HF) between electrocardiography and ballistocardiography in a group of 20 subjects aged between 20 and 40, encompassing both male and female participants with varying body mass index (BMI) categories, including underweight, normal, and obese individuals.

## II. DATA ACQUISITION SYSTEM

The acquisition of electrocardiogram (ECG) and ballistocardiogram (BCG) signals was performed employing the subsequent techniques:

### A. Electrocardiogram

For ECG signal acquisition, a 3-lead ECG acquisition system [10] was utilized. This system encompassed a bio-amplifier which pre-amplified the signal to increase signal strength above noise level.

### B. Ballistocardiogram

Two piezoelectric sensors were connected serially to obtain the BCG signal. These sensors were mounted on a compact single-sided circuit board. The data acquisition (DAQ) card [11], [12], [13] from National Instruments 9201 was employed for signal acquisition. The NI 9201 has a 12-bit resolution and features 8 analog inputs. Additionally, it supports Wi-Fi connectivity. In order to facilitate wireless communication with a laptop, the wireless NI CDAQ 9191 was connected to the DAQ. The establishment of wireless connections between the DAQ and the laptop was achieved using NI Measurement Automation Explorer (MAX).

## III. PREPROCESSING

Subsequently, the acquired ECG and BCG signals were preprocessed using MATLAB. The preprocessing steps involved amplification and denoising.

**ECG:** The signal from the amplifier was passed through a bandpass filter with cut-off frequencies set at 5 Hz and 20 Hz [14]. This step effectively eliminated noise and baseline wandering.

**BCG:** The BCG signal obtained had an amplitude ranging from 30-70 mV, which was subsequently amplified utilizing an instrumentation amplifier with a gain of 10 [15]. The amplified signal was then subjected to band-pass filtering to acquire a signal within the range of 0.1-30 Hz. This filtering step also eliminated baseline wandering.

## IV. HEART RATE CALCULATION

The heart rate (HR) was calculated using the ECG and BCG signals, employing the following methods:

ECG: The QRS complexes present in the ECG signal were detected through the utilization of the Pan-Tomkins algorithm [14]. The locations of the peaks and their corresponding amplitudes were subsequently determined. Instantaneous RR intervals as well as the mean RR interval were computed. Finally, the HR was determined based on the mean RR interval.

BCG: The peaks in the BCG signal were detected utilizing the BCG Heart Beat Detection Method [15]. The locations of the peaks and their corresponding amplitudes were determined accordingly. Instantaneous JJ intervals and the mean JJ interval were computed. The HR was then calculated based on the mean JJ interval.

## V. HEART RATE VARIABILITY ANALYSIS

Heart rate variability (HRV) represents an invaluable non-invasive tool for assessing cardiac autonomic activity [2]. Typically, HRV is computed from beat-to-beat (RR) interval series derived from electrocardiograms (ECGs). Optimal quality ECG/BCG recordings serve as the foundation for accurate HRV analysis [16].

In ECG, the RR intervals indicate the variation between consecutive heartbeats. The analysis of heart rate variability (HRV) measurements revolves around the study of these RR intervals and how they change over time. The MATLAB program was used to determine both time domain and frequency domain indices of HRV. In the time domain, values for SDNN, RMSSD, and pNN50 were calculated. In the frequency domain, VLF, LF, and HF were also determined [17], [18].

In BCG, JJ intervals can be used as an alternative to RR intervals in ECG. The time and frequency domain indices of HRV were evaluated in a similar manner to the ECG analysis.

## VI. RESULTS

### A. Electrocardiography

1) **ECG Signal Preprocessing:** The acquired ECG signal underwent a series of preprocessing steps, including bandpass filtering and averaging using a moving window integrator. The derivative of the integrated signal was then computed, followed by squaring the resultant signal. The outcomes of these preprocessing steps are depicted in Figure 1.

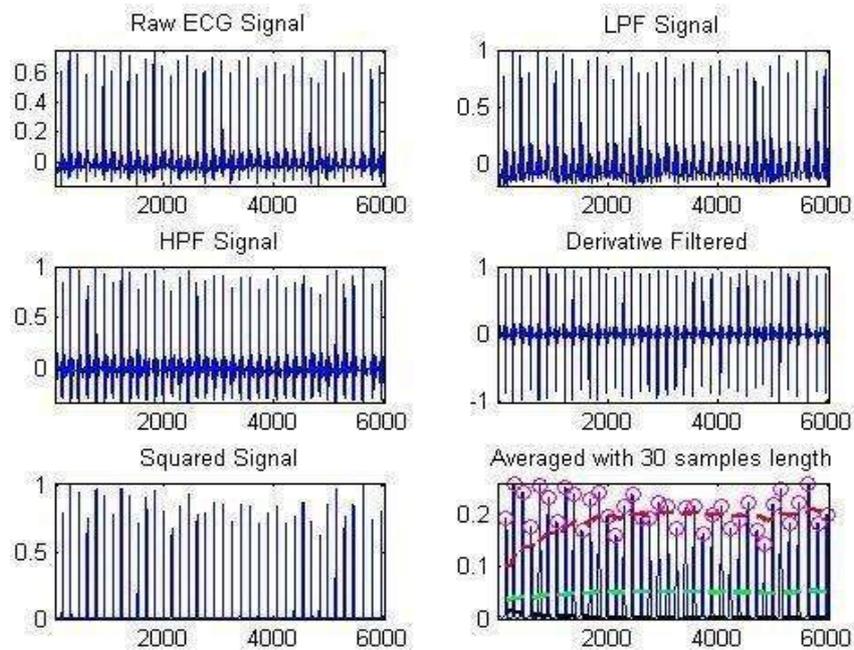

**Figure 1.** ECG Signal Preprocessing

2) **QRS Complex Detection:** QRS complexes were detected utilizing the bandpass-filtered and integrated signal. This process is illustrated in Figure 2.

3) **Heart Rate:** The variations in heart rate are displayed in Figure 3.

The time domain HRV indices are as follows,

| | |
|---|---|
| Mean Heart Rate (in bpm): | 74.924 |
| Standard Deviation of all normal RR Intervals (SDNN in ms): | 133.851 |
| Root Mean Square Successive Differences (RMSSD in ms): | 39.197 |
| Percentage of differences between successive NN intervals that are greater than 50ms (percentage units): | 9.737 |

4) **RR Intervals:** The RR intervals are presented in Figure 4.

5) **Frequency Domain HRV Indices:** The Power Spectral Density of the ECG signal is shown in Figure 5.

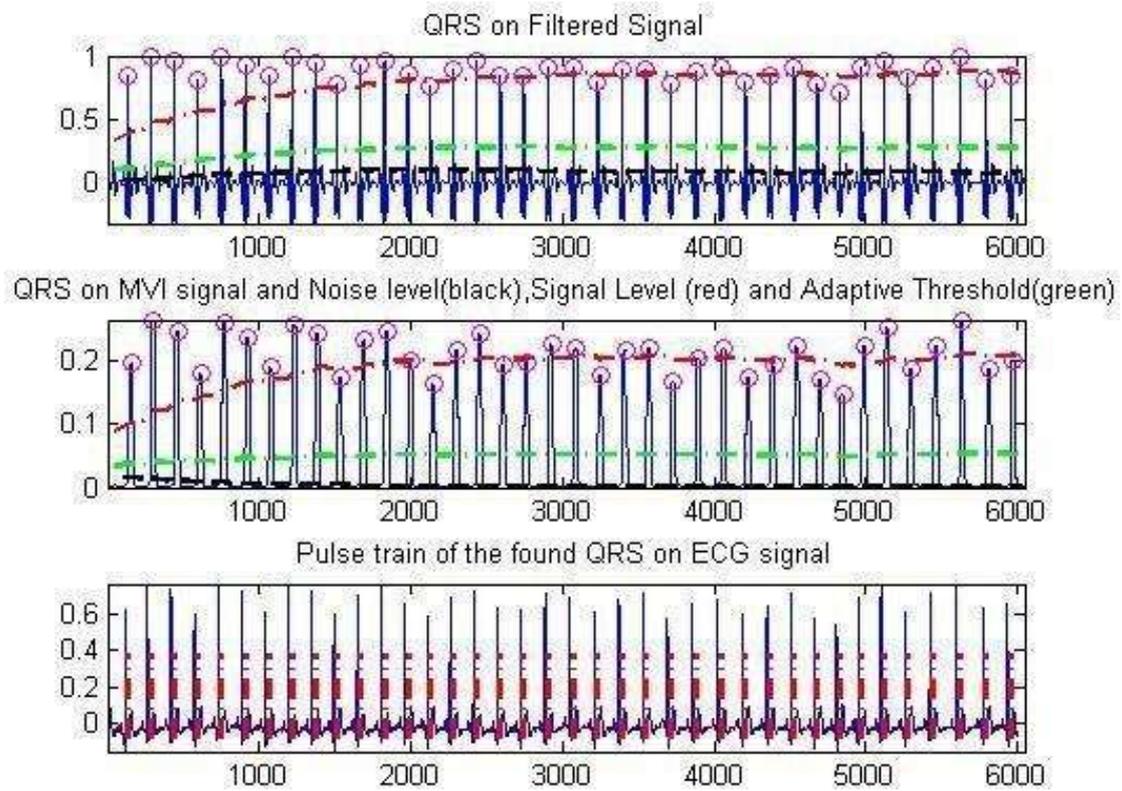

**Figure 2.** Detection of QRS Complexes

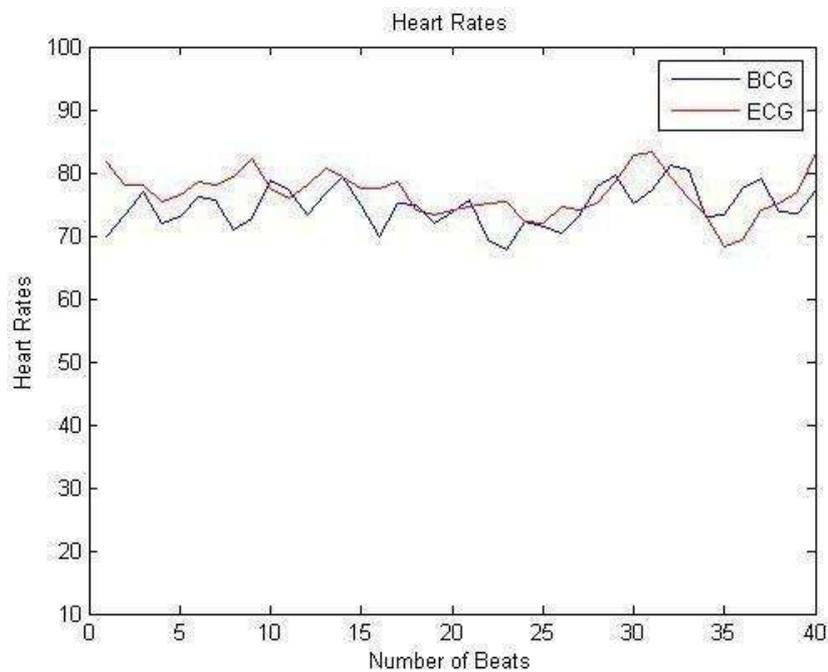

**Figure 3.** Heart Rate variation

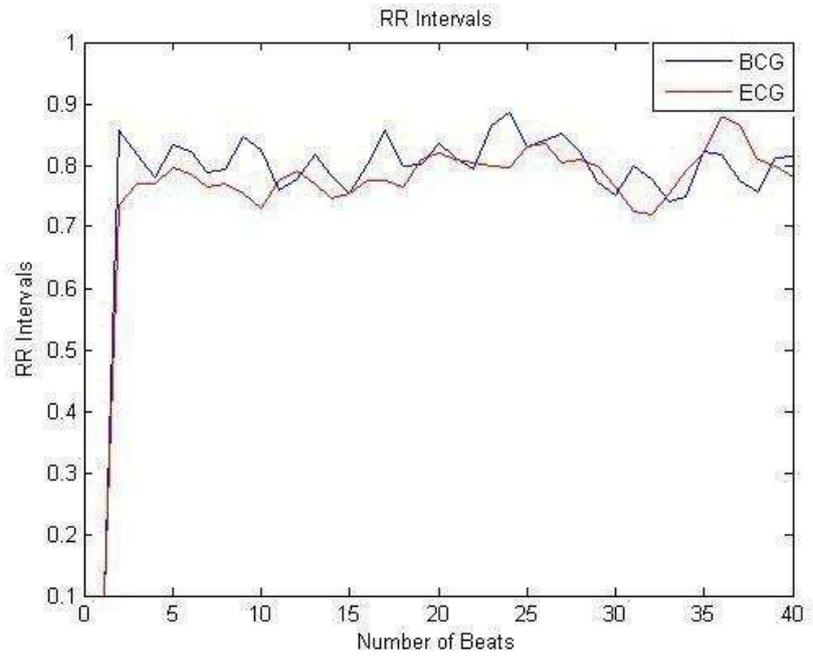

**Figure 4.** RR Intervals

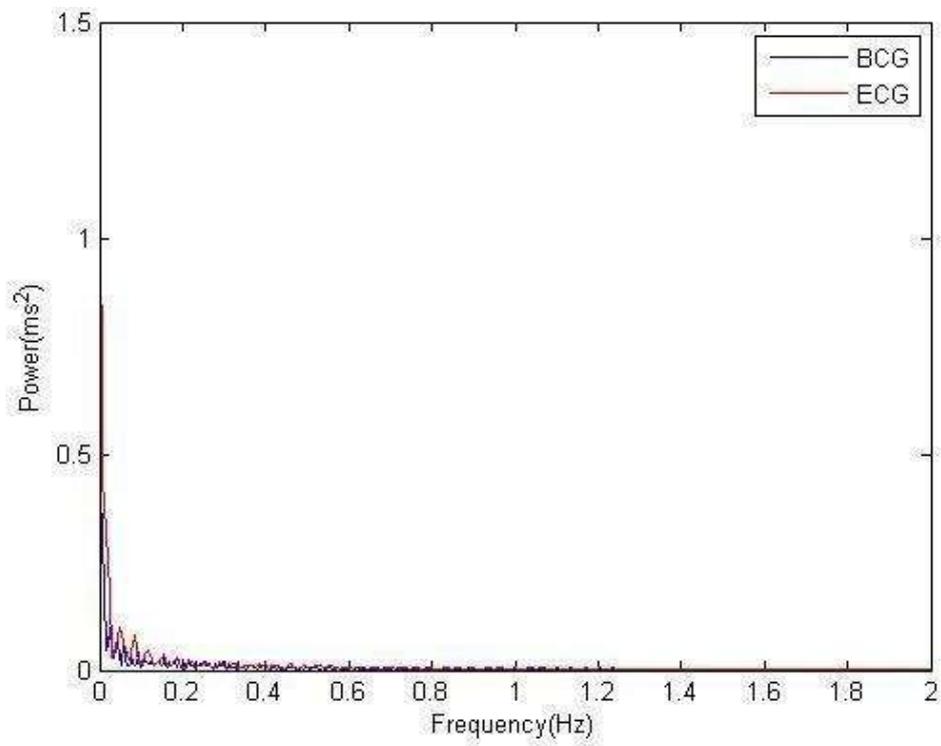

**Figure 5:** Power Spectral Density of the BCG and ECG Signals

### B. Ballistocardiography

1) **BCG Signal Preprocessing:** Similar to ECG, the acquired BCG signal was preprocessed through bandpass filtering and averaging using a moving window integrator. The derivative of the integrated signal was computed, and subsequently, the square of the derivative was taken. The outcomes of these preprocessing steps are presented in Figure 6.

2) **J Peak Detection:** The J peaks were detected using the bandpass-filtered and integrated BCG signal, as illustrated in Figure 7.

3) **Heart Rate:** The variations in heart rate are shown in Figure 7.

   The Time Domain HRV Indices are as follows:

   | | |
   |---|---|
   | Mean Heart Rate (in bpm): | 74.673 |
   | Standard Deviation of all normal JJ Intervals (SDNN in ms): | 132.568 |
   | Root Mean Square Successive Differences (RMSSD in ms): | 40.19 |
   | Percentage of differences between successive NN intervals that are greater than 50ms (percentage units): | 9.756 |

4) **JJ Intervals:** The JJ intervals are depicted in Figure 7.

5) **Frequency Domain HRV Indices:** The Power Spectral Density of the BCG signal is displayed in Figure 5.

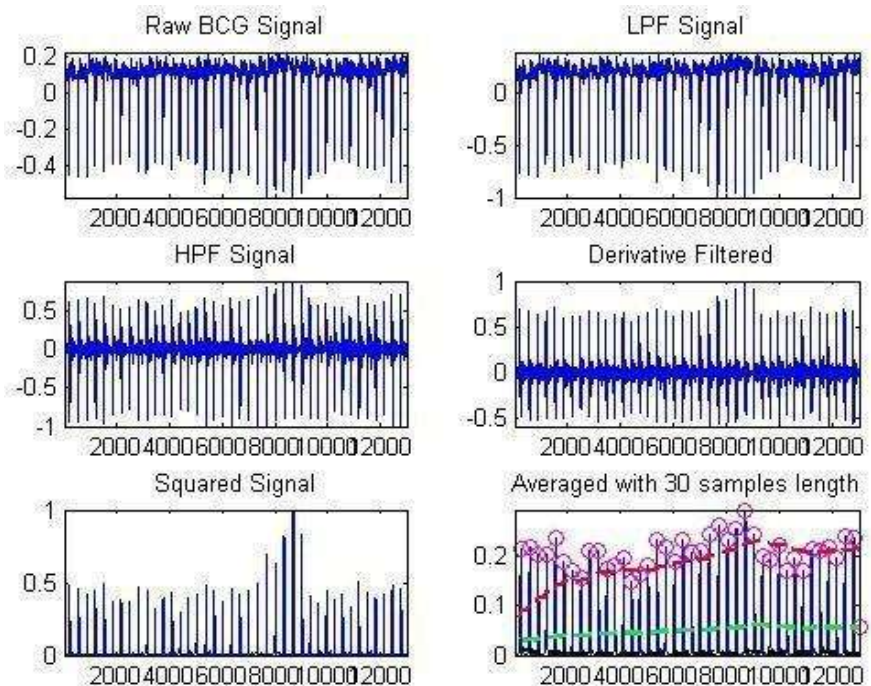

**Figure 6:** BCG Signal Preprocessing

### C. Comparison of HRV Indices of ECG and BCG

On comparing the HRV Indices it is understood that the Ballistocardiogram signals can also provide the same performance as that of the Electrocardiogram signals. The HRV Indices (Time domain and Frequency domain) of both signals are highly correlated. For a single-subject case, the SDNN obtained from ECG was 133.85 ms; the same parameter obtained from BCG was 132.57. As we go for the case of RMSSD it is 39.197 for ECG and 40.19 for BCG, the pNN50 from ECG was 9.7368 and from BCG was 9.7561. So it is clear that the time domain HRV indices obtained from ECG and BCG are highly correlated. The results presented in Figures 5 demonstrate that the similarity between ECG and BCG-derived HRV indices extends to the frequency domain as well. This finding was consistent across 20 subjects, leading us to conclude that long-term monitoring with BCG is indeed a feasible alternative.

VII.     CONCLUSION

The goal of this paper was to compare the Heart Rate Variability (HRV) indices of Electrocardiogram (ECG) and Ballistocardiogram (BCG). After observing the HRV indices of ECG and BCG in 20 subjects, we concluded that there is a high correlation between the HRV (time domain and frequency domain) indices of ECG and BCG. We also found that BCG can be more comfortable for the subject than ECG during data acquisition. BCG uses only sensors that can be laid down under the bed for data collection, whereas ECG uses electrodes that have to be attached to the body of the subject. This difference in data acquisition methods can disturb the subject's freedom of movement for long-term monitoring cases. Since BCG acquisition assures the subject more freedom of movement and since the HRV indices of ECG and BCG are the same, long-term monitoring with BCG is indeed a feasible alternative that ensures a good performance.